\documentstyle{article}

\input amssymb.sty

\renewcommand{\thefootnote}{\fnsymbol{footnote}}
\newfont{\gothic}{eufm10 scaled\magstep1}
\newcommand{\eg}{\mbox{{\it e.g.} }}
\newcommand{\ie}{\mbox{{\it i.e.} }}
\newcommand{\complex}{{\Bbb C}}
\newcommand{\integer}{{\Bbb Z}}
\newcommand{\comm}[2]{\left[#1, #2\right]}
\newcommand{\acomm}[2]{\left\{#1, #2\right\} }
\newcommand{\half}{\frac{1}{2}}
\newcommand{\inv}[1]{\frac{1}{#1}}
\newcommand{\pder}{\partial}
\newcommand{\W}{{\cal W}}
\newcommand{\M}{{\cal M}}

\newcommand{\var}[1]{\delta_2\left(#1\right)}
\newcommand{\conf}[1]{\var{\epsilon_{#1}}}

\newcommand{\algthree}[1]{\delta_3\left({#1}\right)}
\newcommand{\algfour}[1]{\delta_4\left({#1}\right)}
\newcommand{\meas}{\int_{\Sigma}{\rm d}^2 z\,}
\newcommand{\anom}[3]{\Delta^{(#1)}_{#2}\left[#3\right]}
\newcommand{\coup}{\left(\frac{12}{c}\right)}
\newcommand{\wthree}{W_3}
\newcommand{\wfour}{W_4}
\newcommand{\poiss}[3]{\acomm{#1}{#2}_{#3}}
\newcommand{\g}{\gothic g}
\newcommand{\cder}{\bar{\rm D}}

\begin{document}
\begin{titlepage}
\begin{center}
\noindent 09 February 1999\hfill DIAS-STP-98-11\\
\hfill {\tt hep-th/9809078}

\vspace{2cm}

{\large \bf Ward Identities and Anomalies in Pure $\W_4$ Gravity}

\vspace{1 cm}

\setcounter{footnote}{0}
\renewcommand{\thefootnote}{\arabic{footnote}}

{\bf Paul Watts}\footnote{E-mail: {\it watts@stp.dias.ie}, Tel:
+353-1-6140148, Fax: +353-1-6680561, Home Page: {\it
http://www.stp.dias.ie/~watts}}

\vspace{1 cm}

Dublin Institute for Advanced Studies\\
School of Theoretical Physics\\
10 Burlington Road\\
Dublin 4\\\
Ireland

\vspace{2 cm}

{\bf Abstract}

\end{center}

\noindent $\W_4$ gravity is treated algebraically, represented by a set of
transformations on classical fields.  The Ward identities of the theory are
determined by requiring the algebra to close.  The general forms for the
anomalies are found by looking for solutions to the Wess-Zumino consistency
conditions, and some specific cases are considered.

\bigskip

\noindent PACS-95: 11.25.Hf 11.30.Ly\\
\noindent MSC-91: 81T40 81T50

\bigskip


\end{titlepage}
\newpage
\setcounter{page}{1}
\renewcommand{\thepage}{\arabic{page}}

\section{Introduction}
\setcounter{equation}{0}

Since conformal symmetries play such a large role in modern physics, it may
be of some interest to see to what degree such symmetries may be extended.
One such extension to consider is the role of $\W$-algebras in
2-dimensional physics \cite{Zam}.  A $\W_N$ algebra is, simply put, any
consistent algebra which contains the Virasoro algebra, as well as a tower
of Virasoro-primary currents such that $N$ is the highest conformal weight.
In general, for $N\in\integer$, this requires that there are also primary
fields with weights $3,4,\dots,N$.  Thus, we have not only the usual spin 2
field which appears in 2-dimensional conformal field theory, but also
fields with spins up to $N$.  What makes $\W$-algebras particularly notable
is that the commutation relations are generally {\em not} linear; the
commutator between two generators can be any polynomial function of the
basis elements, rather than just a linear combination of generators.  The
resulting algebra is thus no longer Lie but rather Poisson.  (For a
comprehensive overview of the general subject, see \cite{BS}).

The minimal extension of the Virasoro algebra in this manner is the $\W_3$
algebra, which has been treated extensively in far too many works to fully
catalogue here (see \cite{BMP} and references therein, although
\cite{OSSvN} deserves special mention as a major influence on the present
work, both in notation and approach).  In this case, the nonlinearity of
the theory shows up in the fact that the generators of the Virasoro algebra
appear not only linearly, but quadratically.  $\W_4$ is the first instance
in which not only do the Virasoro generators appear cubically, but the
additional $\W_3$ currents themselves show up quadratically.  The structure
of this particular algebra (and $\W$ algebras in general) has been examined
both from BRST \cite{Z,Hornfeck} and OPE \cite{BFK,Zhu} points of view.

$\W_N$ gravity \cite{Hull1,Pope} is the gauged version of a $\W_N$ theory:
Additional fields, one for each current, are introduced such that the
$\W_N$ algebra is preserved even when the gauge symmetry (\ie coordinate
transformations) is included.  However, as in regular conformal theories,
this results in the appearance of gravitational Ward identities and
anomalies.  Most of the work treating these in a $\W$-algebra context have
been for $\W_3$ gravity, and fall into three basic categories: Actually
finding the effective action, Ward identities and anomalies arising from a
$\W$-symmetric quantum field theory \cite{SSvN,OSSvN,deBoer}; explicitly
constructing a free-field realisation of the algebra \cite{Hull2}; or
examining the BRST algebra \cite{GGL,AAC}.  When the anomalies arising from
these approaches are compared, it is not always obvious they are the same.
However, each result turns out to be a specific case of a more general
expression for the anomalies of the theory, one obtained purely on
algebraic grounds, as shown (for the first and last approaches) in
\cite{Watts}.

\bigskip

In this work we attempt to apply the algebraic approach just cited to
$\W_4$: First, we introduce the set of transformations on classical fields
which gives a representation of the algebra, and then extend the theory to
$\W_4$ gravity by gauging the original $\W_4$ theory.  In doing so, we find
that conditions have to be placed on the fields to close the algebra, which
we call `Ward identities', despite the fact that our theory is a purely
algebraic one, with no {\it a priori} connection to an effective field
theory or the like.  In fact, it is for this very reason that we call the
theory `pure $\W_4$ gravity'.  The form of these Ward identities suggests a
basis for the anomalies, and after checking which particular combinations
of these basis elements fulfill the Wess-Zumino consistency conditions, the
general forms are found.  Finally, we find the expressions for the
anomalies for several specific examples, and comment on their possible
connections with known theories.

\section{$\W_4$ Algebra}
\setcounter{equation}{0}

We start with a 2-dimensional space $\Sigma$ with coordinates $z$ and
$\bar{z}$ (and corresponding derivatives $\pder$ and $\bar{\pder}$); the
classical fields which appear in $\W_4$ are the familiar energy-momentum
tensor $T(z,\bar{z})$ and the two $\W$-currents, $\wthree(z,\bar{z})$ and
$\wfour(z,\bar{z})$.  By definition, the $\W_4$ transformations must
include the conformal transformation $\delta_2$, which on $T$ has the form
\begin{equation}
\conf{}T=\left(\frac{c}{12}\right)\pder^3\epsilon+\epsilon\pder
T+2\pder\epsilon T,
\end{equation}
where $c$ is the central charge and $\epsilon(z,\bar{z})$ the conformal
variation parameter.  This gives a representation of the Virasoro algebra
(which, in an abuse of terminology, will occasionally be referred to as the
`$\W_2$ algebra'):
\begin{equation}
\comm{\conf{1}}{\conf{2}}=\var{\poiss{\epsilon_1}{\epsilon_2}{2}},
\end{equation}
with the Poisson bracket between the $\W_2$ variation parameters defined
as
\begin{equation}
\poiss{\epsilon_1}{\epsilon_2}{2}:=\pder\epsilon_1\epsilon_2-\epsilon_1
\pder\epsilon_2.\label{poiss1}
\end{equation}

A `primary field' in this language is a field $\phi_h(z,\bar{z})$ which
transforms as
\begin{equation}
\conf{}\phi_h=\epsilon\pder\phi_h+h\pder\epsilon\phi_h,\label{primary}
\end{equation}
and what we'd like to do is to find a way to introduce $\wthree$ and
$\wfour$ and their $\W_3$ and $\W_4$ transformation laws in a way such that
they are Virasoro primary, $\W_2$ is a subalgebra, and the entire algebra
closes.  The first and second of these criteria are easily fulfilled by
saying
\begin{eqnarray}
\conf{}\wthree&=&\epsilon\pder\wthree +3\pder\epsilon\wthree,\nonumber\\
\conf{}\wfour&=&\epsilon\pder\wfour+4\pder\epsilon\wfour.
\end{eqnarray}

The last criterion can be satisfied uniquely as well (otherwise this would
be a very short paper): First, the $\W_3$ transformations are given by
\begin{eqnarray}
\algthree{\lambda}T&=&2\lambda\pder\wthree+3\pder\lambda\wthree,
\nonumber\\
\algthree{\lambda}\wthree&=&\left(\frac{c}{12}\right)\pder^5\lambda+2
\lambda\pder^3T+9\pder\lambda\pder^2T+15\pder^2\lambda\pder T+10\pder^3
\lambda T\nonumber\\
&&+\lambda\pder\wfour+2\pder\lambda\wfour+\coup\left(16\lambda T\pder
T+16\pder\lambda T^2\right),\nonumber\\
\algthree{\lambda}\wfour&=&\lambda\pder^3\wthree+6\pder\lambda\pder^2
\wthree+14\pder^2\lambda\pder\wthree+14\pder^3\lambda\wthree\nonumber\\
&&+\coup\left(18\lambda T\pder\wthree+25\lambda\pder T\wthree+52\pder
\lambda T\wthree\right).
\end{eqnarray}
The algebra of these transformations does not close even if we include the
conformal transformations, because unlike $\W_2$, $\W_3$ is not a
subalgebra of $\W_4$, so `$\W_3$ transformations' is something of a
misnomer.  To complete the algebra, we have to include the $\W_4$
transformations
\begin{eqnarray}
\algfour{\xi}T&:=&3\xi\pder \wfour+4\pder\xi\wfour,\nonumber\\
\algfour{\xi}\wthree&:=&5\xi\pder^3\wthree+20\pder\xi\pder^2\wthree+28
\pder^2\xi\pder \wthree+14\pder^3\xi \wthree\nonumber\\
&&+\coup\left(34\xi T\pder\wthree+27\xi\wthree\pder T+52\pder\xi T
\wthree\right),\nonumber\\
\algfour{\xi}\wfour&:=&\left(\frac{c}{12}\right)\pder^7\xi+3\xi\pder^5
T+20\pder\xi\pder^4T+56\pder^2\xi\pder^3T+84\pder^3\xi\pder^2T\nonumber\\
&&+70\pder^4\xi\pder T+28\pder^5\xi T-\xi\pder^3\wfour-5\pder\xi\pder^2
\wfour-9\pder^2\xi\pder\wfour-6\pder^3\xi\wfour\nonumber\\
&&+\coup\left[\xi\left(177\pder T\pder^2T+78T\pder^3T\right)+\pder\xi
\left(352T\pder^2T+295(\pder T)^2\right)\right.\nonumber\\
&&+588\pder^2\xi T\pder T+196\pder^3\xi T^2-14\xi T\pder\wfour-14\xi\pder
T\wfour-28\pder\xi T\wfour\nonumber\\
&&\left.+75\xi\wthree\pder\wthree+75\pder\xi\wthree^2\right]\nonumber\\
&&+\coup^2\left(432\xi T^2\pder T+288\pder\xi T^3\right).
\end{eqnarray}
Not surprisingly, the conformal transformations take the $\W$
transformations into themselves:
\begin{eqnarray}
\comm{\conf{}}{\algthree{\lambda}}&=&\algthree{\poiss{\epsilon}{
\lambda}{3}},\nonumber\\
\comm{\conf{}}{\algfour{\xi}}&=&\algfour{\poiss{\epsilon}{\xi}{4}},
\end{eqnarray}
where
\begin{eqnarray}
\poiss{\epsilon}{\lambda}{3}&:=&2\pder\epsilon\lambda-\epsilon\pder\lambda,
\nonumber\\
\poiss{\epsilon}{\xi}{4}&:=&3\pder\epsilon\xi-\epsilon\pder\xi.
\label{poiss2}
\end{eqnarray}

At this point, it may be useful to make a comment on the notation:
Throughout this work, $\W_2$, $\W_3$ and $\W_4$ transformation parameters
will always be referred to by $\epsilon$, $\lambda$ and $\xi$ respectively.
Furthermore, the subscripts on the Poisson brackets will indicate what type
of variation parameter the resulting bracket is, \eg
$\poiss{\epsilon}{\xi}{4}$ takes a $\W_2$ parameter and a $\W_4$ parameter
and spits back a $\W_4$ parameter.

That being said, the rest of the algebra and the other Poisson brackets can
now be written:  The commutators between the transformations are
\begin{eqnarray}
\comm{\algthree{\lambda_1}}{\algthree{\lambda_2}}&=&\var{\poiss{
\lambda_1}{\lambda_2}{2}}+\algfour{\poiss{\lambda_1}{\lambda_2}{4}},
\nonumber\\
\comm{\algthree{\lambda}}{\algfour{\xi}}&=&\var{\poiss{\lambda}{\xi}{2}}
+\algthree{\poiss{\lambda}{\xi}{3}},\nonumber\\
\comm{\algfour{\xi_1}}{\algfour{\xi_2}}&=&\var{\poiss{\xi_1}{\xi_2}{2}}
+\algthree{\poiss{\xi_1}{\xi_2}{3}}+\algfour{\poiss{\xi_1}{\xi_2}{4}},
\end{eqnarray}
with the Poisson brackets
\begin{eqnarray}
\poiss{\lambda_1}{\lambda_2}{2}&:=&2\lambda_2\pder^3\lambda_1-
3\pder\lambda_2\pder^2\lambda_1+3\pder^2\lambda_2\pder \lambda_1-2\pder^3
\lambda_2 \lambda_1\nonumber\\
&&+\coup 16\left(\lambda_2\pder\lambda_1-\pder\lambda_2\lambda_1\right)
T,\nonumber\\
\poiss{\lambda}{\xi}{2}&:=&\coup\left(27\pder\lambda\xi\wthree-25\lambda
\pder\xi\wthree-7\lambda\xi\pder\wthree\right),\nonumber\\
\poiss{\xi_1}{\xi_2}{2}&:=&3\xi_2\pder^5\xi_1-5\pder\xi_2\pder^4\xi_1+6
\pder^2\xi_2\pder^3\xi_1-6\pder^3\xi_2\pder^2\xi_1+5\pder^4\xi_2\pder
\xi_1-3\pder^5\xi_2 \xi_1\nonumber\\
&&+\coup\left[57\left(\xi_2\pder \xi_1-\pder\xi_2\xi_1\right)\pder^2T+57
\left(\xi_2\pder^2\xi_1-\pder^2\xi_2 \xi_1\right)\pder T\right.\nonumber\\
&&+\left(78\xi_2\pder^3\xi_1-118\pder\xi_2\pder^2\xi_1+118\pder^2
\xi_2\pder\xi_1-78\pder^2\xi_2\xi_1\right)T\nonumber\\
&&\left.-14\left(\xi_2\pder\xi_1-\pder\xi_2\xi_1\right)\wfour\right]+\coup^2
432(\xi_2\pder\xi_1-\pder\xi_2\xi_1)T^2,\nonumber\\
\poiss{\lambda}{\xi}{3}&:=&5\xi\pder^3\lambda-5\pder\xi\pder^2\lambda+3
\pder^2\xi\pder\lambda-\pder^3\xi\lambda\nonumber\\
&&+\coup\left(34\xi\pder\lambda T-18\pder\xi\lambda T+7\xi\lambda\pder
T\right),\nonumber\\
\poiss{\xi_1}{\xi_2}{3}&:=&\coup 75\left(\xi_2\pder\xi_1-\pder\xi_2\xi_1
\right)\wthree,\nonumber\\
\poiss{\lambda_1}{\lambda_2}{4}&:=&\lambda_2\pder\lambda_1-\pder\lambda_2
\lambda_1,\nonumber\\
\poiss{\xi_1}{\xi_2}{4}&:=&-\xi_2\pder^3\xi_1+2\pder\xi_2\pder^2\xi_1-
2\pder^2\xi_2\pder\xi_1+\pder^3\xi_2\xi_1\nonumber\\
&&-\coup 14\left(\xi_2\pder\xi_1-\pder\xi_2\xi_1\right)T.\label{poiss3}
\end{eqnarray}

Therefore, we get a three-dimensional representation of the full $\W_4$
algebra on $T$, $\wthree$ and $\wfour$.  (In fact, since $\W_4$ itself has
three generators, this is just the adjoint representation.)

\section{$\W_4$ Gravity and Ward Identities}
\setcounter{equation}{0}

We now want to introduce gravity by gauging the $\W_4$ theory just
discussed.  We call it `gravity' because we interpret our three
$\W$-transformations as arising from coordinate transformations, so we need
to include three `metrics' $h$, $g_3$ and $g_4$, each changing
inhomogeneously under the $\W_{2,3,4}$ variations respectively, but still
giving a representation of $\W_4$.  However, up until now, we have said
very little about the geometrical nature of our transformations, merely
defining them on a set of classical fields and then blindly manipulating
them algebraically.  We can no longer use this approach if we want to talk
about gravity, and must now be more specific.

As we did above, we start with the Virasoro algebra: Recall that a
conformal transformation is a coordinate transformation on $\Sigma$ under
which the invariant length d$s^2=h_{ab}(x)$d$x^a$d$x^b$ rescales by an
overall factor, where $x^{1,2}$ are the coordinates and $h_{ab}$ the metric
on $\Sigma$.  The definition of an object $\phi_h(x)$ of conformal weight
$h$ (not to be confused with the metric) in this picture is one which under
a conformal transformation $x\mapsto x'$ satisfies
\begin{equation}
\phi_h'(x')=\det{}^h\left(\frac{\pder x}{\pder x'}\right)\phi_h(x),
\end{equation}
which, for an infinitesimal transformation $x':=x-\epsilon(x)$, becomes
\begin{equation}
\conf{}\phi_h=\epsilon^a\pder_a\phi_h+h\pder_a\epsilon^a\phi_h.
\end{equation}
Looking at our original definition (\ref{primary}), it is straightforward
to see that in the chosen coordinate system $\left(z,\bar{z}\right)$, the
transformations are simply $z\mapsto z-\epsilon\left(z,\bar{z}\right)$ and
$\bar{z}\mapsto\bar{z}$.  In order for this transformation to truly be
conformal, we need to choose the light-cone gauge, in which case the metric
takes the form d$s^2:=$d$z$d$\bar{z}+h\left(z,\bar{z}\right)$d$\bar{z}^2$
for some quantity $h\left(z,\bar{z}\right)$ \cite{P}.  In this gauge,
d$s^2\mapsto\left(1-\pder\epsilon\right)$d$s^2$ provided that $h$
transforms according to
\begin{equation}
\conf{}h=\bar{\pder}\epsilon+\epsilon\pder h-h\pder\epsilon.\label{conf-h}
\end{equation}
This fits right in with the criteria we'd like for the gauge field:  It is
primary (of weight -1) except for the inhomogeneous first term, so we take
it as the field to include in the formulation of $\W_4$ gravity.

Now, on to the other two `metrics', $g_3$ and $g_4$: Note that
(\ref{conf-h}) can also be written as $\bar{\pder}\epsilon+\poiss{h}{
\epsilon}{2}$.  The appearance of the Poisson bracket in a linear and local
way, plus the fact that we would like $g_{3,4}$ (like $W_{3,4}$) to be
primary, motivates the choices
\begin{eqnarray}
\conf{}g_3&:=&\poiss{g_3}{\epsilon}{3}\nonumber\\
&=&\epsilon\pder g_3-2\pder\epsilon g_3,\nonumber\\
\conf{}g_4&:=&\poiss{g_4}{\epsilon}{4}\nonumber\\
&=&\epsilon\pder g_4-3\pder\epsilon g_4,
\end{eqnarray}
so $g_{3,4}$ are primary of weight $-2$ and $-3$, and therefore we will
still have a representation of the Virasoro algebra.

Continuing along this train of thought, we construct the $\W_3$ and $\W_4$
transformations of $h$, $g_3$ and $g_4$ with the following requirements:
First, in analogy to $\conf{}h$, $\algthree{\lambda}g_3$ and
$\algfour{\xi}g_4$ must include a $\bar{\pder}\lambda$ and $\bar{\pder}\xi$
respectively; and secondly, all other pieces of the variations must be
expressible purely in terms of the Poisson brackets, but only linearly and
locally.  These criteria, together with the ever-present demand that $\W_2$
remains a subalgebra, lead to
\begin{eqnarray}
\algthree{\lambda}{h}&:=&\poiss{g_3}{\lambda}{2}+\poiss{g_4}{\lambda}{2},
\nonumber\\
\algthree{\lambda}{g_3}&:=&\bar{\pder}\lambda+\poiss{h}{\lambda}{3}+
\poiss{g_4}{\lambda}{3},\nonumber\\
\algthree{\lambda}{g_4}&:=&\poiss{g_3}{\lambda}{4},\nonumber\\
\algfour{\xi}h&:=&\poiss{g_3}{\xi}{2}+\poiss{g_4}{\xi}{2},\nonumber\\
\algfour{\xi}g_3&:=&\poiss{g_3}{\xi}{3}+\poiss{g_4}{\xi}{3},\nonumber\\
\algfour{\xi}g_4&:=&\bar{\pder}\xi+\poiss{h}{\xi}{4}+\poiss{g_4}{\xi}{4},
\end{eqnarray}
or, more explicitly,
\begin{eqnarray}
\algthree{\lambda}h&=&2\lambda\pder^3g_3-3\pder\lambda\pder^2g_3+3
\pder^2\lambda\pder g_3-2\pder^3\lambda g_3\nonumber\\
&&+\coup\left(16\lambda T\pder g_3-16\pder\lambda Tg_3+25\lambda\wthree
\pder g_4+7\lambda\pder\wthree g_4-27\pder\lambda\wthree g_4\right),
\nonumber\\
\algthree{\lambda}g_3&=&\bar{\pder}\lambda+2\lambda\pder h-\pder\lambda
h+\lambda\pder^3 g_4-3\pder\lambda\pder^2g_4+5\pder^2\lambda\pder
g_4-5\pder^3 \lambda g_4\nonumber\\
&&+\coup\left(18\lambda T\pder g_4-7\lambda\pder Tg_4-34\pder\lambda
Tg_4\right),\nonumber\\
\algthree{\lambda}g_4&=&\lambda\pder g_3-\pder\lambda g_3,
\end{eqnarray}
and
\begin{eqnarray}
\algfour{\xi}h&=&3\xi\pder^5g_4-5\pder\xi\pder^4g_4+6\pder^2\xi\pder^3g_4
-6\pder^3\xi\pder^2g_4+5\pder^4\xi\pder g_4-3\pder^5\xi g_4\nonumber\\
&&+\coup\left[57\left(\xi\pder g_4-\pder\xi g_4\right)\pder^2T+57\left(
\xi\pder^2g_4-\pder^2\xi g_4\right)\pder
T\right.\nonumber\\
&&+\left(78\xi\pder^3g_4-118\pder\xi\pder^2g_4+118\pder^2\xi\pder g_4-
78\pder^2\xi g_4\right)T\nonumber\\
&&\left.+27\xi\wthree\pder g_3-25\pder\xi\wthree g_3-7\xi\pder\wthree
g_3-14\left(\xi\pder g_4-\pder\xi g_4\right)\wfour\right]\nonumber\\
&&+\coup^2 432(\xi\pder g_4-\pder\xi g_4)T^2,\nonumber\\
\algfour{\xi}g_3&=&5\xi\pder^3g_3-5\pder\xi\pder^2g_3+3\pder^2\xi\pder
g_3-\pder^3\xi g_3\nonumber\\
&&+\coup\left[34\xi T\pder g_3-18\pder\xi Tg_3+7\xi\pder Tg_3+75\left(
\xi\pder g_4-\pder\xi g_4\right)\wthree\right],\nonumber\\
\algfour{\xi}g_4&=&\bar{\pder}\xi+2\xi\pder h-\pder\xi h-\xi\pder^3
g_4+2\pder\xi\pder^2 g_4-2\pder^2\xi\pder g_4+\pder^3\xi g_4\nonumber\\
&&-\coup 14\left(\xi\pder g_4-\pder\xi g_4\right)T.
\end{eqnarray}

Unfortunately, now the algebra no longer closes; this is a result of the
fact that not all the Poisson brackets respect $\bar{\pder}$, \ie $\bar{
\pder}\poiss{\alpha}{\beta}{i}-\poiss{\bar{\pder}\alpha}{\beta}{i}-
\poiss{\alpha}{\bar{\pder}\beta}{i}$ does not always vanish ($\alpha$ and
$\beta$ are arbitrary transformation parameters).  For instance,
\begin{eqnarray}
\comm{\algthree{\lambda_1}}{\algthree{\lambda_2}}h&=&\left(\var{\poiss{
\lambda_1}{\lambda_2}{2}}+\algfour{\poiss{\lambda_1}{\lambda_2}{4}}\right)
h\nonumber\\
&&-\coup 16\left(\lambda_2\pder\lambda_1-\pder\lambda_2\lambda_1\right)
\omega_2,
\end{eqnarray}
where $\omega_2$ is the quantity
\begin{equation}
\omega_2:=\bar{\pder}T-\left(\frac{c}{12}\right)\pder^3h-h\pder T-2\pder
hT-2g_3\pder\wthree-3\pder g_3\wthree-3g_4\pder\wfour-4\pder g_4\wfour.
\label{ward1}
\end{equation}
Given our assumptions as to the forms of the variations, the above
transformations seem to be the only choices, and there would appear to be a
problem.  Luckily, there is a resolution: Notice that if we define the
operator $\bar{nabla}$ as
\begin{equation}
\bar{\nabla}:=\bar{\pder}-\var{h}-\algthree{g_3}-\algfour{g_4},
\end{equation}
then $\omega_2\equiv\bar{\nabla}T$.  We can therefore define two new quantities
$\omega_{3,4}$ as $\bar{\nabla}W_{3,4}$:
\begin{eqnarray}
\omega_3&:=&\bar{\pder}\wthree-\left(\frac{c}{12}\right)\pder^5g_3-h\pder
\wthree-3\pder h\wthree-2g_3\pder^3T-9\pder g_3\pder^2T-15\pder^2g_3\pder 
T\nonumber\\
&&-10\pder^3g_3T-g_3\pder\wfour-2\pder g_3\wfour-5 g_4\pder^3\wthree-20
\pder g_4\pder^2\wthree-28\pder^2 g_4\pder\wthree\nonumber\\
&&-14\pder^3g_4\wthree-\coup\left(16g_3T\pder T+16\pder g_3T^2+34g_4T
\pder\wthree+27g_4\pder T\wthree\right.\nonumber\\
&&\left.+52\pder g_4T\wthree\right),\nonumber\\
\omega_4&:=&\bar{\pder}\wfour-\left(\frac{c}{12}\right)\pder^7g_4-h\pder
\wfour-4\pder h\wfour-g_3\pder^3\wthree-6\pder g_3\pder^2\wthree-14\pder^2
g_3\pder\wthree\nonumber\\
&&-14\pder^3g_3\wthree-3g_4\pder^5T-20\pder g_4\pder^4T-56\pder^2g_4\pder^3
T-84\pder^3g_4\pder^2T-70\pder^4g_4\pder T\nonumber\\
&&-28\pder^5g_4T+g_4\pder^3\wfour+5\pder g_4\pder^2\wfour+9\pder^2g_4\pder
\wfour+6\pder^3g_4\wfour\nonumber\\
&&-\coup\left[18g_3T\pder\wthree+25g_3\pder T\wthree+52\pder g_3T\wthree
+g_4\left(177\pder T\pder^2T\right.\right.\nonumber\\
&&\left.+78T\pder^3T\right)+\pder g_4\left(352T\pder^2T+295(\pder
T)^2\right)+588\pder^2g_4T\pder T+196\pder^3g_4T^2\nonumber\\
&&\left.-14g_4T\pder\wfour-14 g_4\pder T\wfour-28\pder g_4T\wfour+75g_4
\wthree\pder\wthree+75\pder g_4\wthree^2\right]\nonumber\\
&&-\coup^2\left(432g_4T^2\pder T+288\pder g_4T^3\right).
\label{ward2}
\end{eqnarray}
The usefulness of introducing these quantities lies in the fact that they
change into one another under the $\W_4$ transformations, \eg
\begin{eqnarray}
\algthree{\lambda}\omega_4&=&\lambda\pder^3\omega_3+6\pder\lambda\pder^2
\omega_3+14\pder^2\lambda\pder\omega_3+14\pder^3\lambda\omega_3\nonumber\\
&&+\coup\left(18\lambda\omega_2\pder\wthree+25\lambda\pder\omega_2\wthree
+52\pder\lambda\omega_2\wthree\right.\nonumber\\
&&\left.+18\lambda T\pder\omega_3+25\lambda\pder T\omega_3+52\pder\lambda
T\omega_3\right).
\end{eqnarray}
Hence, since $\omega_{2,3,4}$ span an invariant subspace of the space of
fields, we can mod them out, \ie take them to vanish identically.  This
means that all the commutation relations of the $\W_4$ transformations now
are satisfied on {\em all} of the fields, and the algebra closes.

Algebraically, we should have no qualms in setting the $\omega$s to zero,
but what's the interpretation of this requirement from a physics point of
view?  Simply that the vanishing of these quantities give the Ward
identities (WIs) of the physical theory: We can think of the classical
fields we deal with here as resulting from the quantisation of a $\W_4$
invariant quantum field theory, and that the WIs are necessary to ensure
that the $\W_4$ algebra is preserved at the level of the effective theory,
precisely as we have just seen.  So in reality, the vanishing of
(\ref{ward1}) and (\ref{ward2}) implies that we have only three physical
degrees of freedom rather than six.  Since the $\omega$s span an invariant
subspace, setting them equal to zero amounts to imposing a set of
first-class constraints on our theory.

(For an example of why one can think of the vanishing of the $\omega$s as
equivalent to the usual WIs, we can refer back to \cite{OSSvN}, where the
authors start with an explicitly $\W_3$-symmetric theory and show that as a
consequence of this symmetry, the effective currents and metrics must
satisfy two differential equations, which are precisely the same as our
conditions $\omega_{2,3}\equiv 0$ with $T\rightarrow\left(\frac{12}{c}
\right)u$, $W_3\rightarrow-\left(\frac{12}{c}\right)\inv{\sqrt{30}}v$,
$h\rightarrow h$ and $g_3\rightarrow-\inv{\sqrt{30}}b$ (the weight 4 fields
do not appear in $\W_3$ gravity, of course).)

\bigskip

As a slight digression, we notice that a geometrical interpretation can
also be provided: If $\M$ is the space of functions on $\Sigma$, then
$\bar{\nabla}$ is just $\bar{\pder}$ on elements of $\M$.  Furthermore, the
definitions of the transformation laws of the metrics automatically imply
that they are all annihilated by $\bar{\nabla}$, and therefore
$\bar{\nabla} \omega_{2,3,4}$ are each linear in the $\omega$s; thus, we
can consistently impose $\omega_{2,3,4}\equiv 0$, \ie that $\bar{\nabla}$
vanishes on $T$ and $W_{3,4}$.  This means that we can define the nilpotent
exterior derivative $\cder:={\rm d}\bar{z}\,\bar{\nabla}$, and immediately
see that $H^0(\M,\complex;\cder)$ is the space of holomorphic functions on
$\Sigma$ and that all physical fields (multiplied by d$\bar{z}$) must
belong to $H^1(\M,\complex;\cder)$ (with, in the latter case, the
equivalence relation $\cder\alpha\simeq 0$, $\alpha\in\M$ ensuring that our
currents are not simply functions on $\Sigma$).

\section{Anomalies}
\setcounter{equation}{0}

\subsection{General $\W_4$ Anomalies}

The purpose of this subsection is to try to find the most general possible
forms of the anomalies in $\W_4$ gravity.  As with the rest of this work,
this will be done in an algebraic manner: Suppose we have an algebra
represented on classical fields $\{\phi_i\}$ by means of a transformation
rule $\phi_i\mapsto\phi_i+\delta(\alpha)\phi_i$, such that $\comm{\delta
(\alpha_1)}{\delta(\alpha_2)}=\delta\left(\acomm{\alpha_1}{\alpha_2}
\right)$ for some Poisson bracket $\{\cdot,\cdot\}$.  In this context, an
anomaly $\Delta[\alpha]$ is a functional of the variation parameter of the
form
\begin{equation}
\Delta[\alpha]\equiv\meas\alpha A\left(\{\phi\}\right)
\end{equation}
which satisfies the Wess-Zumino consistency condition (WZCC)
\begin {equation}
\delta\left(\alpha_1\right)\Delta\left[\alpha_2\right]-\delta\left(
\alpha_2\right)\Delta\left[\alpha_1\right]=\Delta\left[\acomm{\alpha_1}{
\alpha_2}\right].\label{wzcc}
\end{equation}

The WZCC is satisfied iff the algebra closes, of course.  However, if there
is an invariant subspace with basis $\{\omega_m\}$ such that the
commutation relations have the form
\begin{equation}
\left(\comm{\delta\left(\alpha_1\right)}{\delta\left(\alpha_2\right)}-
\delta\left(\acomm{\alpha_1}{\alpha_2}\right)\right)\phi_i=\sum_mF_{im}
\left(\alpha_1,\alpha_2;\{\phi\}\right)\omega_m,
\end{equation}
then the right-hand side of (\ref{wzcc}) will pick up a term whose
integrand is a linear combination of the $\omega$s.  If we then close the
algebra by modding out by this subspace, this extraneous term will vanish,
and the WZCC will work after all.  Thus, in looking for the most general
possible form of the anomaly, we must allow for such terms to pop up when
checking the WZCC.

To try to find generic expressions for the $\W_4$ anomalies, we propose to
start with a basis of functionals and see which particular combinations
satisfy the nine WZCCs.  However, as just argued, we should only wait until
all the computations are done before imposing the WIs, so we allow the
$\omega$s to show up in the integrands of our proposed anomalies.  Now,
notice that $\coup$ serves throughout all our discussions as a sort of
expansion parameter, and therefore it follows that the anomalies themselves
are expressible as a sum of terms of various powers of $\coup$.
Furthermore, unlike $\pder$, $\bar{\pder}$ never appears with more than
unit degree in any of the transformation laws.  With all this in mind, and
looking at the expressions (\ref{ward1}) and (\ref{ward2}) for the
$\omega$s, we propose the following basis for the $\W_2$ anomalies:
\begin{eqnarray}
\anom{1}{2}{\epsilon}&:=&\left(\frac{c}{12}\right)\meas\epsilon\pder^3
h,\nonumber\\
\anom{2}{2}{\epsilon}&:=&\meas\epsilon\bar{\pder}T,\nonumber\\
\anom{3}{2}{\epsilon}&:=&\meas\epsilon\left(h\pder T+2\pder hT+2g_3\pder
\wthree+3\pder g_3\wthree\right.\nonumber\\
&&\left.+3g_4\pder\wfour+4\pder g_4\wfour\right);
\label{anom2}
\end{eqnarray}
for the $\W_3$ anomalies:
\begin{eqnarray}
\anom{1}{3}{\lambda}&:=&\left(\frac{c}{12}\right)\meas\lambda\pder^5g_3,
\nonumber\\
\anom{2}{3}{\lambda}&:=&\meas\lambda\bar{\pder}\wthree,\nonumber\\
\anom{3}{3}{\lambda}&:=&\meas\lambda\left(h\pder\wthree+3\pder h\wthree
+2g_3\pder^3T+9\pder g_3\pder^2T+15\pder^2g_3\pder T+10\pder^3
g_3T\right.\nonumber\\
&&\left.+g_3\pder\wfour+2\pder g_3\wfour+5 g_4\pder^3\wthree+20\pder
g_4\pder^2\wthree+28\pder^2 g_4\pder\wthree+14\pder^3g_4\wthree\right),
\nonumber\\
\anom{4}{3}{\lambda}&:=&\coup\meas\lambda\left(16g_3T\pder T+16\pder g_3
T^2+34g_4T\pder\wthree\right.\nonumber\\
&&\left.+27g_4\pder T\wthree+52\pder g_4T\wthree\right);
\label{anom3}
\end{eqnarray}
and finally for the $\W_4$ anomalies:
\begin{eqnarray}
\anom{1}{4}{\xi}&:=&\left(\frac{c}{12}\right)\meas\xi\pder^7g_4,
\nonumber\\
\anom{2}{4}{\xi}&:=&\meas\xi\bar{\pder}\wfour,\nonumber\\
\anom{3}{4}{\xi}&:=&\meas\xi\left(h\pder\wfour+4\pder h\wfour+g_3\pder^3
\wthree+6\pder g_3\pder^2\wthree+14\pder^2g_3\pder\wthree+14\pder^3g_3
\wthree\right.\nonumber\\
&&+3 g_4\pder^5T+20\pder g_4\pder^4T+56\pder^2g_4\pder^3T+84\pder^3
g_4\pder^2T+70\pder^4g_4\pder T+28\pder^5g_4T\nonumber\\
&&\left.-g_4\pder^3\wfour-5\pder g_4\pder^2\wfour-9\pder^2g_4\pder\wfour
-6\pder^3g_4\wfour\right),\nonumber\\
\anom{4}{4}{\xi}&:=&\coup\meas\xi\left[18g_3T\pder\wthree+25g_3\pder
T\wthree+52\pder g_3T\wthree+g_4\left(177\pder T\pder^2T+78T\pder^3T
\right)\right.\nonumber\\
&&+\pder g_4\left(352T\pder^2T+295(\pder T)^2\right)+588\pder^2
g_4T\pder T+196\pder^3g_4T^2\nonumber\\
&&\left.-14g_4T\pder\wfour-14 g_4\pder T\wfour-28\pder g_4T\wfour+75g_4
\wthree\pder\wthree+75\pder g_4\wthree^2\right],\nonumber\\
\anom{5}{4}{\xi}&:=&\coup^2\meas\xi\left(432g_4T^2\pder T+288\pder g_4T^3
\right).\label{anom4}
\end{eqnarray}
This is obviously not the only basis possible, and for the purposes of our
discussion it's more convenient to introduce the six functionals
$\delta_{2,3,4}L$ and $\Omega_{2,3,4}$ defined by
\begin{eqnarray}
\delta_2L[\epsilon]&:=&-\anom{1}{2}{\epsilon}-\anom{2}{2}{\epsilon},
\nonumber\\
\delta_3L[\lambda]&:=&-\anom{1}{3}{\lambda}-\anom{2}{3}{\lambda}+
\anom{4}{3}{\lambda},\nonumber\\
\delta_4L[\xi]&:=&-\anom{1}{4}{\xi}-\anom{2}{4}{\xi}+\anom{4}{4}{\xi}
+2\anom{5}{4}{\xi},\nonumber\\
\Omega_2[\epsilon]&:=&\anom{2}{2}{\epsilon}-\anom{1}{2}{\epsilon}-
\anom{3}{2}{\epsilon},\nonumber\\
\Omega_3[\lambda]&:=&\anom{2}{3}{\lambda}-\anom{1}{3}{\lambda}-
\anom{3}{3}{\lambda}-\anom{4}{3}{\lambda},\nonumber\\
\Omega_4[\xi]&:=&\anom{2}{4}{\xi}-\anom{1}{4}{\xi}-\anom{3}{4}{\xi}-
\anom{4}{4}{\xi}-\anom{5}{4}{\xi}.
\end{eqnarray}
These choices are not arbitrary; the first three are, respectively, the
$\W_2$, $\W_3$ and $\W_4$ variations of the quantity
\begin{equation}
L:=\meas\left(Th+\wthree g_3+\wfour g_4\right).\label{legendre}
\end{equation}
The other three are simply the integrals of the products between the
appropriate $\omega$s and parameters, \eg $\Omega_2[\epsilon]\equiv\meas
\epsilon\omega_2$.

To show why these are useful, we stick these into the WZCCs.  First of all,
a straightforward computation shows that
\begin{eqnarray}
\conf{1}\Omega_2\left[\epsilon_2\right]&=&\Omega_2\left[\poiss{\epsilon_1}{
\epsilon_2}{2}\right],\nonumber\\
\var{\epsilon}\Omega_3[\lambda]&=&\Omega_3\left[\poiss{\epsilon}{\lambda}{
3}\right],\nonumber\\
\var{\epsilon}\Omega_4[\xi]&=&\Omega_4\left[\poiss{\epsilon}{\xi}{4}\right].
\end{eqnarray}
Similar computations show that the variations of the $\Omega$s parallel the
algebra itself; in other words, if we write the algebra symbolically as
$\comm{\delta_i}{\delta_j}=f_{ij}{}^k\delta_k$, where $f_{ij}{}^k$ gives
the structure `constants' (actually Poisson brackets) of $\W_4$, then we
find $\delta_i\Omega_j=f_{ij}{}^k\Omega_k$.

The same holds true for the $\delta L$s, namely, they respect the
commutation relations for the algebra, except for the fact that when a
Poisson bracket has explicit field dependence, an additional $\Omega$ term
is picked up.  For example,
\begin{eqnarray}
&\algthree{\lambda_1}\delta_3L\left[\lambda_2\right]-\algthree{
\lambda_2}\delta_2L\left[\lambda_1\right]-\delta_2L\left[\poiss{\lambda_1}{
\lambda_2}{2}\right]-\delta_4L\left[\poiss{\lambda_1}{\lambda_2}{4}\right]
=&\nonumber\\
&\Omega_2\left[\frac{\pder}{\pder T}\poiss{\lambda_1}{\lambda_2}{2}
\right],&\label{anomQ}
\end{eqnarray}
where the $T$ derivative will just pick out the $O\left(\frac{12}{c}
\right)$ piece of $\poiss{\lambda_1}{\lambda_2}{2}$.  Of course, when the
$\omega$s are modded out, these $\Omega$ terms vanish.

So these six functionals automatically satisfy the WZCCs when the WIs are
imposed, and therefore we can eliminate two each of (\ref{anom2}),
(\ref{anom3}) and (\ref{anom4}) in favour of them.  We choose
$\Delta^{(2)}_i$ and $\Delta^{(3)}_i$, because the former are the only ones
with $\bar{\pder}$s, and the latter the most complicated.

Now, to finish finding the most general anomalies, we have to check the
WZCCs for each of the twelve proposed basis anomalies.  This involves some
lengthy computations which we will not include here; suffice it to say that
we find that apart from the $\Omega$s and $\delta L$s, only three other
specific combinations of the $\Delta$s work:
\begin{eqnarray}
\anom{0}{2}{\epsilon}&:=&\anom{1}{2}{\epsilon},\nonumber\\
\anom{0}{3}{\lambda}&:=& \anom{1}{3}{\lambda}-\anom{4}{3}{\lambda},
\nonumber\\
\anom{0}{4}{\xi}&:=&\anom{1}{4}{\xi}-\anom{4}{4}{\xi}-2\anom{5}{4}{\xi}.
\end{eqnarray}
These form a set satisfying the WZCCs, in exactly the same manner as the
$\delta L$s do, except with the opposite sign in front of the $\Omega$, \ie
(\ref{anomQ}) with $\delta_iL\rightarrow\Delta^{(0)}_i$ and $\Omega_i
\rightarrow-\Omega_i$.

So, finally, we conclude that the most general possible forms for the
$\W_2$, $\W_3$ and $\W_4$ anomalies satisfying the WZCCs, modulo the Ward
identities, are
\begin{eqnarray}
\Delta_2[\epsilon]&=&a\anom{0}{2}{\epsilon}+b\delta_2L[\epsilon]+r_2
\Omega_2[\epsilon],\nonumber\\
\Delta_3[\lambda]&=&a\anom{0}{3}{\lambda}+b\delta_3L[\lambda]+r_3\Omega_3[
\lambda],\nonumber\\
\Delta_4[\xi]&=&a\anom{0}{4}{\xi}+b\delta_4L[\xi]+r_4\Omega_4[\xi],
\end{eqnarray}
where $a$, $b$ and $r_{2,3,4}$ are constants.

\subsection{`Ward-Free' Anomalies}

At this point, we've done all we can algebraically.  To say anything more
about the anomalies requires that we put further conditions on what
constitutes an `acceptable anomaly' in addition to the WZCCs.  For
instance, if we demand that we never have to invoke the WIs at all, then
that leads to the unique choice of $a=b$ and $r_{2,3,4}=0$.  This follows
from the facts that the $\Delta^{(0)}_i$s and the $\delta_iL$s violate the
WZCCs by the `opposite amounts' of the $\Omega$s, and because the $\Omega$s
themselves satisfy the WZCCs iff they are put to zero afterward.
Therefore, by taking the sum of the former ($a=b$) and eliminating the
latter ($r_{2,3,4}=0$), the anomalies obtained,
\begin{eqnarray}
\Delta_2[\epsilon]=\meas\epsilon\bar{\pder}T,&\Delta_3[\lambda]=\meas
\lambda\bar{\pder}\wthree,&\Delta_4[\xi]=\meas\xi\bar{\pder}\wfour,
\end{eqnarray}
satisfy the WZCCs.  So even though we must demand that the $\omega$s vanish
to close the algebra, there do in fact exist anomalies which respect the
WZCCs without invoking this condition.

\subsection{BRST Anomalies}

The fact that we had to impose a set of first-class constraints to close
the $\W_4$ gravity algebra ($\omega_{2,3,4}\equiv 0$) suggests that an
interesting case to look at might be where we actually have a BRST algebra:
This amounts to introducing three fermionic ghost fields $b_{2,3,4}$ and a
new operator $Q$, the BRST charge, defined by
\begin{equation}
Q:=\var{b_2}+\algthree{b_3}+\algfour{b_4}.
\end{equation}
$Q^2$ vanishes identically on $T$, $\wthree$ and $\wfour$ provided $Q$ acts
on the ghosts as
\begin{eqnarray}
Qb_2&:=&-\half\poiss{b_2}{b_2}{2}+\half\poiss{b_3}{b_3}{2}+\half\poiss{
b_4}{b_4}{2}-\poiss{b_3}{b_4}{2},\nonumber\\
Qb_3&:=&-\poiss{b_2}{b_3}{3}+\poiss{b_3}{b_4}{3}-\poiss{b_4}{b_4}{3},
\nonumber\\
Qb_4&:=&-\poiss{b_2}{b_4}{4}+\half\poiss{b_3}{b_3}{4}+\half\poiss{b_4}{
b_4}{4},
\end{eqnarray}
where the Poisson brackets are those obtained when $b_{2,3,4}$ are treated
as $\W_{2,3,4}$ variation parameters respectively.  These definitions also
lead to the vanishing of $Q^2$ on the ghosts themselves as well.  (Note
that since the ghosts are fermionic, the ordering in the above Poisson
brackets is important, so the reader is advised to use (\ref{poiss1}),
(\ref{poiss2}) and (\ref{poiss3}) {\em exactly} as written, \eg
$\poiss{b_2}{b_2}{2}=\pder b_2b_2-b_2\pder b_2=-2b_2\pder b_2$.)

$Q^2$ vanishes on the ghosts as well, but not on the metrics; for example,
\begin{eqnarray}
Q^2h&=&\coup\left[16b_3\pder b_3\omega_2+57b_4\pder b_4\pder^2\omega_2+57
b_4\pder^2b_4\pder\omega_2+\left(78b_4\pder^3b_4\right.\right.\nonumber\\
&&\left.\left.-118\pder b_4\pder^2b_4\right)\omega_2-27\pder b_3b_4
\omega_3+25b_3\pder b_4\omega_3+7b_3b_4\pder\omega_3-14b_4\pder
b_4\omega_4\right]\nonumber\\
&&+\coup^2 864b_4\pder b_4T\omega_2.
\end{eqnarray}
This should come as no surprise at all, because the nilpotency of the BRST
charge is dependent upon the fact that the structure constants of the
symmetry algebra satisfy the Jacobi identity, which is {\em not} true for a
$\W$-algebra, due to the field dependence of the Poisson brackets.  The
condition for the satisfaction of the Jacobi identities, \ie the closure of
the algebra, is merely the imposition of the WIs, as we have seen before,
so $Q$ is nilpotent iff the $\omega$s vanish.  Or, if we reverse the
argument, we could have introduced the BRST transformations and imposed
nilpotency of $Q$ to find the WIs.

To find the BRST anomalies, we look for functionals of our fields with unit
ghost number which are $Q$-closed.  Such functionals must be integrals
where the integrand is linear in one of the $b$s, so it makes sense to look
at our basis of anomalies where the arguments are replaced by the ghost
fields.  It is a straightforward exercise to show
\begin{eqnarray}
Q\left(\delta_2L\left[b_2\right]+\delta_3L\left[b_3\right]+\delta_4L\left[
b_4\right]\right)&=&0,\nonumber\\
Q\left(\Omega_2\left[b_2\right]+\Omega_3\left[b_3\right]+\Omega_4\left[
b_4\right]\right)&=&-\Omega_2\left[Qb_2\right]-\Omega_3\left[Qb_3\right]-
\Omega_4\left[Qb_4\right],\nonumber\\
Q\left(\anom{0}{2}{b_2}+\anom{0}{3}{b_3}+\anom{0}{4}{b_4}\right)&=&0,
\end{eqnarray}
so upon imposing the WIs, all of these fit the bill, so the most general
BRST anomaly has the form
\begin{equation}
\Delta=\Delta_1\left[b_1\right]+\Delta_2\left[b_2\right]+\Delta_3\left[b_3
\right],
\end{equation}
for arbitrary $a$, $b$ and $r_{2,3,4}$.

\subsection{Anomalies from Effective Action}

Another case we might want to consider is where we think of our anomalies
as arising from an effective $\W_4$ symmetric field theory, in which case
the anomalies will be the variations of the effective action.  To be
precise, if we start with a theory of $\W_4$ gravity with metrics $h'$,
$g_3'$ and $g_4'$ and invariant action $S'$, then by introducing currents
$T$, $\wthree$ and $\wfour$ we find the generating partition function
\begin{equation}
Z:=\int\left[{\rm d}h'\right]\left[{\rm d}g_3'\right]\left[{\rm d}g_4'
\right]e^{iS'+i\meas\left(Th'+\wthree g_3'+\wfour g_4'\right)}.
\end{equation}
The effective metrics $h$, $g_3$ and $g_4$ are, respectively, the
functional derivatives of $-i\ln Z$ with respect to $T$, $\wthree$ and
$\wfour$, and so the effective action is obtained via the usual Legendre
transformation $S:=-i\ln Z-L$, where $L$ is the same functional from
(\ref{legendre}).  Now, if we compute the variation of $S$ under a $\W_4$
transformation, the definitions of the effective metrics immediately lead
to
\begin{equation}
\delta_iS=-\meas\left(T\delta_ih+\wthree\delta_ig_3+\wfour\delta_ig_4
\right).
\end{equation}
If the variations are put in explicitly, we get the result that $\delta_iS
=\Delta^{(0)}_i+\Omega_i$.  However, recall that this is all in the context
of the currents being functions of the effective metrics, which is simply
the statement that the WIs are satisfied.  Thus, a $\W_4$ symmetric field
theory has $a=1$, $b=r_{2,3,4}=0$:
\begin{eqnarray}
\var{\epsilon}S&=&\meas\left(\frac{c}{12}\right)\epsilon\pder^3h,
\nonumber\\
\algthree{\lambda}S&=&\meas\lambda\left[\left(\frac{c}{12}\right)\pder^5
g_3\right.\nonumber\\
&&\left.-\coup\left(16g_3T\pder T+16\pder g_3T^2+34g_4T\pder\wthree+27g_4
\pder T\wthree+52\pder g_4T\wthree\right)\right],\nonumber\\
\algfour{\xi}S&:=&\meas\xi\left\{\left(\frac{c}{12}\right)\xi\pder^7g_4
\right.\nonumber\\
&&-\coup\left[18g_3T\pder\wthree+25g_3\pder T\wthree+52\pder g_3T
\wthree+g_4\left(177\pder T\pder^2T+78T\pder^3T\right)\right.\nonumber\\
&&+\pder g_4\left(352T\pder^2T+295(\pder T)^2\right)+588\pder^2g_4T\pder
T+196\pder^3g_4T^2\nonumber\\
&&\left.-14g_4T\pder\wfour-14g_4\pder T\wfour-28\pder g_4T\wfour+75g_4
\wthree\pder\wthree+75\pder g_4\wthree^2\right]\nonumber\\
&&\left.-\coup^2\left(864g_4T^2\pder T+576\pder g_4T^3\right)\right\}.
\end{eqnarray}

\section{Conclusions}
\setcounter{equation}{0}

In this paper we have shown that $\W_4$, represented as a set of
transformations on classical fields, can be gauged into `pure $\W_4$
gravity' provided that the three currents and three metrics satisfy
relations which may be identified with the Ward identities of the theory.
(Alternatively, this may be thought of as saying that we have a theory on a
fibre bundle with base space $\Sigma$, structure group $\W_4$ and fibre
$H^1(\M,\complex;\cder)$.)  The form of these identities suggests a basis
for the anomalies of the theory, from which we can find the particular
linear combinations which satisfy the Wess-Zumino consistency conditions.
Once found, we can then restrict our theory in various ways (\eg as arising
from a BRST algebra) to determine the anomalies specific to those cases.

\subsection{Connection to Toda Theory}

One possible avenue of further exploration might be a deeper examination of
one of the cases for which the anomalies were found, namely, the one in
which our theory arises from an underlying $\W_4$-symmetric quantum field
theory: Recall the usual conformal case, where the WI is just
\begin{equation}
\bar{\pder}T=\left(\frac{c}{12}\right)\pder^3h+h\pder T+2\pder hT.
\end{equation}
This gives a relation between $T$ and $h$, so in reality there is only one
physical degree of freedom, call it $f_2$.  This appears explicitly when
we solve the above WI, via
\begin{eqnarray}
h=\frac{\bar{\pder}f_2}{\pder f_2},&&T=\left(\frac{c}{12}\right)\frac{\pder
f_2\pder^3f_2-\frac{3}{2}\left(\pder^2f_2\right)^2}{(\pder f_2)^2}.
\end{eqnarray}
The conformal variations on $T$ and $h$ become $\conf{}f_2=\epsilon\pder
f_2$, so the anomaly as a functional of $f_2$ is therefore
\begin{eqnarray}
\Delta[\epsilon]&=&\left(\frac{c}{12}\right)\meas\epsilon\pder^3h
\nonumber\\
&=&\left(\frac{c}{12}\right)\meas\frac{\conf{}f_2}{\pder f_2}\pder^3
\left(\frac{\bar{\pder}f_2}{\pder f_2}\right).
\end{eqnarray}
This must be a conformal variation of an effective action, and when we
integrate it, we find
\begin{equation}
S[f_2]=-\left(\frac{c}{24}\right)\meas\frac{\bar{\pder}f_2}{\pder f_2}
\pder^2\ln\pder f_2.\label{action}
\end{equation}
Actually, we can add any conformally invariant term to this, since it will
not effect the anomaly.  The obvious one is simply some multiple of the
volume of $\Sigma$, which, since the determinant of the metric is unity, is
simply $\meas$.

Suppose we now change coordinates to $\sigma^+:=f_2\left(z,\bar{z}\right)$
and $\sigma^-:=\bar{z}$, and define a new field $\phi_2(\sigma):=-\ln\pder
f_2(z(\sigma))$ (so $\conf{}\phi_2=-\pder\epsilon+\epsilon\pder\phi_2$);
the metric is now the conformal one, d$s^2=e^{\phi_2}$d$\sigma^+$d$
\sigma^-$, and the action takes the form
\begin{equation}
S[\phi_2]=\left(\frac{c}{24}\right)\int_{\Sigma}{\rm d}^2\sigma\left(
\pder_+\phi_2\pder_-\phi_2-\Lambda e^{\phi_2}\right)\label{Liouville}
\end{equation}
(where we have chosen the constant multiplying the volume of $\Sigma$ to be
$-\frac{c\Lambda}{24}$).  This is instantly recognisable as the action for
Liouville gravity, which is no great revelation, since both (\ref{action})
and (\ref{Liouville}) are just specific examples of the general
gravitational action
\begin{equation}
S[h]=-\left(\frac{c}{24}\right)\int_{\Sigma}{\rm d}^2x\sqrt{h}\left(R
\inv{\Delta}R+\Lambda\right)
\end{equation}
evaluated in the appropriate coordinate systems.

So how does this connect to $\W_4$ gravity as examined here?  In principle,
we can follow the same technique as just presented, by finding three fields
$f_2$, $f_3$ and $f_4$ such that the currents and metrics may be given as
functions of these when the WIs are solved.  Then the three anomalies could
be written as functionals of the same fields, and hopefully an effective
action leading to them could be found.  Since this involves solving three
coupled nonlinear partial differential equations, it is obviously easier
said than done, but if possible, then an appropriate change of coordinates
$z\rightarrow\sigma$ and field redefinitions $f_i\rightarrow\phi_i$ may
bring the action into a Liouville-like form.

But in principal, this may have already been done: Recall that there
already exists a generalisation of the Liouville theory which exhibits
$\W$-symmetry, namely, Toda theory \cite{BG,BFOFW}, so it seems very likely
that the action obtained would be the one for a particular Toda theory.
Furthermore, since the number of fields in a Toda theory with associated
simple Lie algebra $\g$ is equal to the rank of $\g$, it follows that since
our theory has three degrees of freedom after solving the WIs, we would be
expecting a $A_3$, $B_3$, $C_3$, $D_3$ or $F_3$ Toda theory.  However, the
only one of these containing fields of weights $2$, $3$ {\em and} $4$ is
the $A_3=SU(4)$ case, so presumably this would be the result of the
computation just outlined.  As of this writing, this equivalence has not
yet been shown, but may serve as a basis for future work.

\section*{Acknowledgements}

I would like to thank Richard Grimm of the Centre de Physique Th\'eorique
in Marseille for introducing me to the subject of $\W$-algebras.  I am also
grateful to Jan-Martin Pawlowski, Sreedhar Vinnakota and especially
Lochlainn \'O Raifeartaigh for helpful suggestions and comments.

\end{document}